\documentstyle[12pt]{article}
\def\nz{\ifmmode {I\hskip -3pt N} \else {\hbox {$I\hskip -3pt N$}}\fi}
\def\zz{\ifmmode {Z\hskip -4.8pt Z} \else
       {\hbox {$Z\hskip -4.8pt Z$}}\fi}
\def\qz{\ifmmode {Q\hskip -5.0pt\vrule height6.0pt depth 0pt
       \hskip 6pt} \'else {\hbox
       {$Q\hskip -5.0pt\vrule height6.0pt depth 0pt\hskip 6pt$}}\fi}
\def\rz{\ifmmode {I\hskip -3pt R} \else {\hbox {$I\hskip -3pt R$}}\fi}
\def\cz{\ifmmode {C\hskip -4.8pt\vrule height5.8pt\hskip 6.3pt} \else
       {\hbox {$C\hskip -4.8pt\vrule height5.8pt\hskip 6.3pt$}}\fi}
\def\tz{\ifmmode {T\hskip -4.8pt\vrule height5.8pt\hskip 6.3pt} \else
       {\hbox {$T\hskip -4.8pt\vrule height5.8pt\hskip 6.3pt$}}\fi}

\def\qed{\hbox {\hskip 1pt \vrule width 4pt height 6pt depth 1.5pt
        \hskip 1pt}\\}% cqfd
\def\and {{\rm \; and \;}}

\def\ep{\varepsilon}
\def\R{{\rz}}
\def\Z{{\zz}}
\def\T{{\tz}}

\def\be{\begin{equation}}
\def\ee{\end{equation}}

\def\ds{\displaystyle}
\def\cos{{\rm cos}}
\def\sin{{\rm sin}}

\newcommand {\pa}{\partial}

\makeatletter
\@addtoreset{equation}{section}

\makeatother

\newtheorem{theorem}{Theorem}[section]

\newtheorem{proposition}[theorem]{Proposition}

\begin{document}

\baselineskip=16pt

\title{Absolute Continuity of the Floquet Spectrum  for a Nonlinearly
Forced
Harmonic  Oscillator}
\date{}
\maketitle
%\date{}
\begin{center}{ Sandro Graffi\footnote{e-mail: graffi@dm.unibo.it. Partly
supported
by MURST, National Research Project "Sistemi dinamici" and by Universit\`a
di Bologna,
Funds for Selected Research Topics.}
\\ {\small Dipartimento di Matematica, Universit\`a di Bologna}
\\ {\small Piazza di Porta S.Donato 5, 40127 Bologna, Italy} \\
and \\
Kenji Yajima\footnote{e-mail: yajima@ms.u-tokyo.ac.jp. Partly supported by
the
Grant-in-Aid for Scientific  Research, The Ministry of Education, Science,
Sports and
Culture, Japan
\#11304006} \\
{\small Department of Mathematical Sciences, \ University of Tokyo} \\
{\small 3-8-1 Komaba, Meguro-ku, Tokyo 153-7815, Japan}}
\end{center}

 %\makeindex

\bibliographystyle{plain}
%\maketitle
%
%RESUME
%
\noindent
\begin{abstract} \noindent
We prove that the Floquet spectrum of the time peri\-o\-dic
Schr\"o\-din\-ger equation
${\ds i \frac{\pa u}{\pa t} = -\frac{1}{2}\Delta u + \frac{1}{2}x^2 + 2\ep
(\sin t )x_1 u+\mu
V(x,t), }$ corresponding to a mildly nonlinear resonant forcing, is purely
absolutely continuous
for
$\mu$ suitably small.
\end{abstract}

% %FIN DE RESUME %
\section{Introduction and statement of the result}

It is well known \cite{HLS} that the spectrum of the Floquet operator
of the resonant, linearly forced Harmonic oscillator
\[
i \frac{\pa u}{\pa t} = -\frac{1}{2}\Delta u + \frac{1}{2}x^2 + 2\ep (\sin
t )x_1 u,
\quad x=(x_1, \ldots, x_n) \in \R^n, \ \ \ep>0
\]
is purely absolutely continuous. We show in this paper that the absolute
continuity of the
Floquet spectrum persists under time-periodic perturbations growing no
faster than
linearly at infinity provided the resonance condition still
holds. Thus we consider the time-dependent Schr\"odinger equation
\be
i \frac{\pa u}{\pa t} = -\frac{1}{2}\Delta u + \frac{1}{2}x^2u + 2\ep(\sin
t )x_1 u
+ \mu V(t,x)u
\label{eq-1}
\ee
and suppose that $V(t,x)$ is a real-valued smooth function of $(t,x)$,
$2\pi$-periodic
with respect to $t$, increasing at most linearly as $|x|$ goes to infinity:

\be
\label{ass}
|\pa_x^\alpha V(t,x) | \leq C_\alpha, \quad |\alpha| \geq 1.
\ee
Under this condition Eqn. (\ref{eq-1}) generates a unique unitary
propagator $U(t,s)$
on the Hilbert space $L^2(\R^n)$. The Floquet operator is the one-period
propagator
$U(2\pi, 0)$ and we are interested in the nature of its spectrum. It is
well known that the long
time behaviour of the solutions of  (\ref{eq-1}) can be characterized by
means of the spectral
properties of the  Floquet operator (\cite{KY}). Our main result in this
paper is the following
theorem.

\begin{theorem} Let $V$ be as above. Then, for
${\ds |\mu| < \ep \sup_{t, x} |\pa_{x_1} V(t,x)|}$, the spectrum of the
Floquet operator $U=U(2\pi, 0)$ is purely absolutely continuous.
\end{theorem}

\noindent
{\bf Remark}\par\noindent
The above result can be understood in terms of the classical resonance
phenomenon. If $V=0$ the
motions generated by the classical Hamiltonian
$\displaystyle \frac12(p^2+x^2)+2\ep x_1\sin{t}$
undergo a global  resonance between the proper frequency of the harmonic
motions and the frequency
of the linear forcing term. All initial conditions eventially  diverge to
infinity by oscillations
of linearly increasing amplitude.   The quantum counterpart of this
phenomonon is the absolute
continuity of the  Floquet spectrum\cite{HLS}. One might ask whether this
absolute continuity is
stable under  perturbations which destroy the li\-nearity of the forcing
potential.
Theorem 1.1 establishes the stability under perturbations which make the
forcing a non-linear
one  but do not destroy the globality of the resonance phenomenon because
all initial
conditions still diverge by oscillations to infinity. The globality
property of the
resonance phenomonon seems therefore a necessary condition for the absolute
continuity  of the
Floquet spectrum. It is indeed known (\cite{H}) that  the Schr\"odinger
operators
${\ds -\frac{1}{2}\Delta +\frac12 |x|^\alpha + \frac12\ep x_1\sin{\omega
t}}$,
$\alpha>2$, $\omega \in \R$, whose classical counterparts yield local
nonlinear resonances, have no absolutely continuous part in their Floquet
spectrum.
\vskip12pt\noindent
{\bf Notation} We use the vector notation:  for the multiplication operator
$X_j$ by
the variable $x_j$ and the differential operator
${\ds D_j=\frac{1}{i}\frac{\pa}{\pa x_j}}$, $j=1, \ldots, n$,
we denote $X=(X_1, \ldots, X_n)$ and ${\ds D=(D_1, \ldots, D_n)}$.
For a measurable function $W$ and a set of commuting selfadjoint operators
${\cal H}=({\cal H}_1, \ldots, {\cal H}_n)$, $W({\cal H})$ is the operator
defined via functional calculus. We have the identity
\be
{\cal U}^\ast W({\cal H}) {\cal U} = W({\cal U}^\ast {\cal H} {\cal U})
\label{unitary}
\ee
for any unitary operator ${\cal U}$.

\section{Proof of the Theorem}

It is well known (\cite{Ya}) that the nature of the spectrum of the Floquet

operator $U$ is the same (apart from  multiplicities) as that of the
Floquet
Hamiltonian formally given by
\be
{\cal K}u
=-i \frac{\pa u}{\pa t}  -\frac{1}{2}\Delta u + \frac{1}{2}x^2 u+
2\ep (\sin t )x_1 u + \mu V(t,x)u
\label{defK}
\ee
on the Hilbert space ${\bf K}= L^2(\T) \otimes L^2(\R^n)$, where $\T=
\R/2\pi \Z$ is the
circle. More precisely, if ${\cal K}$ is the generator of the one-parameter
strongly
continuous unitary group ${\cal U}(\sigma)$, $\sigma \in \R$, defined by
\be
({\cal U}(\sigma)u)(t) = U(t, t-\sigma) u(t-\sigma) , u=u(t, \cdot) \in
{\bf K},
\ee
then, ${\cal U}(2\pi)= e^{-i2\pi {\cal K}}$ is unitarily equivalent to
${\bf 1}\otimes U(2\pi, 0)$. We set
\[
{\bf D} \equiv C^\infty (\T, {\cal S}(\R^n)).
\]
It is easy to see that:
\begin{enumerate}
\item The function space ${\bf D}$ is dense in ${\bf K}$.
\item ${\bf D}$ is invariant under the action of the group ${\cal
U}(\sigma)$.
\item For $u \in {\bf D}$, ${\cal K}u$ is given by the right hand side of
(\ref{defK}).
\end{enumerate}
It follows that ${\bf D}$ is a core for ${\cal K}$ (\cite{RS})
and ${\cal K}$ is the closure of the operator defined by (\ref{defK}) on
${\bf D}$.

We introduce four unitary operators ${\cal U}_0 \sim {\cal U}_3$ on ${\bf
K}$ and
successively transform ${\cal K}$ by ${\cal U}_j$ as follows: Write
\[
H_0 = -\frac{1}{2}\Delta  + \frac{1}{2}x^2-\frac{1}{2}
\]
and define
\be
{\cal U}_0u(t, \cdot) =e^{-itH_{0}}u(t, \cdot), \quad u \in {\bf K}.
\label{u-0}
\ee

\begin{proposition} {\rm (1)}\ The operator
${\cal U}_0$ is a well defined unitary operator on ${\bf K}$. \\
{\rm (2)}\ ${\cal U}_0$ maps ${\bf D}$ onto itself. \\
{\rm (3)}\ For $ u \in {\bf D}$,
${\cal K}_1 \equiv {\cal U}_0^\ast {\cal K}{\cal U}_0 $ is given by
\be
{\cal K}_1u=-i \frac{\pa u}{\pa t} + 2\ep \sin t (X_1 \cos t + D_1\sin t)u
+
\mu V(t,X \cos t + D\sin t)u + \frac{u}{2}.
\label{1}
\ee
{\rm (4)}\ ${\bf D}$ is a core of ${\cal K}_1$.
\end{proposition}
{\bf Proof}. \ It is well-known that $\sigma(H_0)=\{0,1, \ldots \}$ and we
have
$e^{-2\pi ni H_0}={\bf 1}$. Hence (\ref{u-0}) defines a unitary operator on

${\bf K}$. We have ${\ds {\cal S}(\R^n)= \cap_{k=1}^\infty D(H_0^k)}$ and
(2) follows.
(3) follows from the identity (\ref{unitary}) and the well-known formulae
\[
e^{itH_{0}} X e^{-itH_{0}}= X\cos t + D \sin t, \quad
e^{itH_{0}} D e^{-itH_{0}}= -X\sin t + D \cos  t.
\]
Since ${\bf D}$ is a core of ${\cal K}$ and ${\cal U}_0$ maps ${\bf D}$
onto itself,
${\bf D}$ is also a core for ${\cal K}_1$. \qed

Note that for any linear function $a X + bD +c$ of $X$ and $D$, and $W$
satisfying
(\ref{ass}), $W(aX+bD+c)$ is  a  pseudo-differential operator with Weyl
symbol
$W(ax+b\xi +c)$ (\cite{Hoe}).

To eliminate the term $2\ep X_1 \sin t \cos t$ from ${\cal K}_1$, we define

\be
{\cal U}_1 u(t,x) = e^{i\ep (\cos 2t)x_1/2} u(t,x).
\ee
It is easy to see that ${\cal U}_1$ maps ${\bf D}$ onto itself and
we have
\[
{\cal U}_1^\ast \left(-i\frac{\pa}{\pa t}\right) {\cal U}_1=
\left(-i\frac{\pa}{\pa t}\right) -\ep (\sin 2t)X_1, \quad
{\cal U}_1^\ast D {\cal U}_1= D + \frac{\ep\cos 2t}{2}{\bf e}_1,
\]
on ${\bf D}$. It follows that ${\cal K}_2 \equiv {\cal U}_1^\ast{\cal
K}_1{\cal U}_1$
is given by the closure of
\be
\begin{array}{l}
\ds {\cal K}_2u = -i \frac{\pa u}{\pa t} +2\ep(\sin^2t)D_1 u + \ep^2(\sin^2
t \cos 2t) u  \\
\ds \qquad \qquad + \mu V(t,X \cos t + \sin t(D+\frac{\ep\cos 2t}{2}{\bf
e}_1))u + \frac{u}{2}
\end{array}
\label{K-2}
\ee
defined on ${\bf D}$. We write $2\ep(\sin^2 t)D_1 = \ep D_1-\ep(\cos
2t)D_1$ in the right side of
(\ref{K-2}).

Next,  to eliminate the term $-\ep(\cos 2t)D_1$, we define
\[
{\cal U}_2u(t,x) = e^{i\ep(\sin 2t)D_1/2} u(t,x) = u(t,x+ \ep (\sin 2t){\bf
e}_1/2).
\]
Then, ${\cal U}_2$ maps ${\bf D}$ onto itself and we have on ${\bf D}$
\[
{\cal U}_2^\ast \left(-i\frac{\pa}{\pa t}\right) {\cal U}_2=
\left(-i\frac{\pa}{\pa t}\right) +\ep (\cos 2t)D_1, \quad
{\cal U}_2^\ast X {\cal U}_2= X - \frac{\ep\sin 2t}{2}{\bf e}_1.
\]
It follows, also with the help of the identity (\ref{unitary}),
that ${\cal K}_3 \equiv {\cal U}_2^\ast{\cal K}_2{\cal U}_2$ is the
closure of the operator given on ${\bf D}$ by
\be
\begin{array}{l}
\ds {\cal K}_3 u = -i \frac{\pa u}{\pa t} + \ep D_1u + \ep^2(\sin^2 t\cos 2
t)u  \\
\ds \qquad \qquad + \mu V(t,X\cos t+ D\sin t-\frac{\ep\sin t}{2}{\bf
e}_1)u+ \frac{u}{2}. \end{array}
\label{K-3}
\ee
Here we also used the obvious identity $\cos 2t \sin t -\cos t \sin 2t =
-\sin t$.
\par
We write now
\[
(\sin^2 t)\cos 2t = \frac{1}{2}\cos 2t -\frac{1}{4} \cos 4t - \frac{1}{4}.
\]
and define
\[
{\cal U}_3 u(t,x) = e^{-i\ep^2 (\sin 2t)/4 + i\ep^2 (\sin 4t)/16} u(t,x).
\]
Again ${\cal U}_3$ maps ${\bf D}$ onto itself and
${\cal L} \equiv {\cal U}_3^\ast {\cal K}_2{\cal U}_3$
is the closure of the operator given on ${\bf D}$ by
\be
\begin{array}{l}
\ds {\cal L}u =-i \frac{\pa u}{\pa t} + \ep D_1u +\frac{(2-\ep^2)u}{4} \\
\ds \qquad \qquad + \mu V\left(t,X\cos t+ D\sin t-\frac{\ep {\bf e}_1 \sin
t}{2}\right)u .
\end{array}
\label{trans}
\ee
Thus, ${\cal K}$ is unitarily equivalent to ${\cal L}$ defined as the
closure
of the  operator with domain ${\bf D}$ and action specified  by the right
side  of (\ref{trans}).

\vspace{0.3cm}
\noindent
{\bf Completion of the proof of the Theorem}. \ We apply Mourre's theory of

conjugate operators (\cite{M}; see also\cite{PSS}). We take the selfadjoint
operator ${\cal A}$
defined by
\[
{\cal A}u(t,x) = x_1 u(t,x)
\]
with obvious domain as the conjugate operator for ${\cal L}$, and verify
the conditions (a-e) of
Definition 1 of \cite{M} are satisfied.
\begin{itemize}
\item[(a)] ${\bf D} \subset D({\cal A}) \cap D({\cal L})$ and hence
$D({\cal A}) \cap D({\cal L})$ is a core of ${\cal L}$.
\item[(b)] It is clear that $e^{i\alpha {\cal A}}= e^{i\alpha X_1}$ maps
${\bf D}$ onto
${\bf D}$ and that, for $u \in {\bf D}$, we have
\[
\begin{array}{l}
\ds e^{-i\alpha{\cal A}} {\cal L}e^{i\alpha {\cal A}}u - {\cal L}u =
\ep\alpha u -
\ds \mu V\left(t,X\cos t+ D\sin t-\frac{\ep {\bf e}_1 \sin t}{2}\right)  \\
\ds \qquad \qquad \qquad \qquad + \mu V\left(t,X\cos t+ D\sin
t-\frac{(\ep-2\alpha) {\bf e}_1 \sin
t}{2}\right).
\end{array}
\]
Since
$V(x)- V(x+ \alpha{\bf e}_1 \sin t)$
is bounded with bounded derivatives, the right hand side extends to a
bounded operator
on ${\bf K}$ and it is continuous with respect to $\alpha$ in the operator
norm
topology.
It follows that $e^{i\alpha {\cal A}}$ maps the domain of ${\cal L}$ into
itself and
$\sup_{|\alpha| \leq 1}\|{\cal L}e^{i\alpha {\cal A}}u\|_{{\bf K}}<\infty$
for
any $u \in D({\cal L})$.
\item[(c)] Let us verify the conditions (c'), (i), (ii), (iii) of
Proposition II.1 of
\cite{M} taking $H={\cal L}$, $A={\cal A}$ and ${\cal S}={\bf D}$. The
verification of
these conditions in turn implies (c). First remark that (i) and (ii) are a
direct consequence
of (a) and (b) above. Moreover for any $u \in {\bf D}$ we have
\be
i[{\cal L}, {\cal A}]u =\ep u + \mu \sin t\cdot \pa_{x_1}V\left(t,X\cos t
+D\sin t
- \frac{\ep\sin t}{2}{\bf e}_1\right)u
\label{comm}
\ee
The right hand side extends to a bounded operator $C$ in ${\bf K}$ which,
following
\cite{M}, we denote $i[{\cal L}, {\cal A}]^\circ$. The boundedness implies
a
fortiori  Condition (iii) and hence (c) is verified.
\item[(d)] By direct computation we have
\be
i[[{\cal L}, {\cal A}]^\circ,{\cal A}]u=\mu \sin^2{t}(\partial^2_{x_1}V)
\left(t,X\cos t +D\sin t - \frac{\ep\sin t}{2}{\bf e}_1\right)u.
\label{above}
\ee
The right hand side extends to a bounded operator on ${\bf K}$. It follows
that
$[{\cal L}, {\cal A}]^\circ D({\cal A}) \subset D({\cal A})$ and
(\ref{above}) holds for
$u \in D({\cal A})$. Hence $[[{\cal L}, {\cal A}]^\circ, {\cal A}]$ defined
on
$D({\cal L}) \cap D({\cal A})$ is bounded and this yields (d).
\item[(e)] The operator norm of
${\ds u \mapsto \sin t\cdot \pa_{x_1}V\left(t,X\cos t +D\sin t
- \frac{\ep\sin t}{2}{\bf e}_1\right)u} $ is bounded by
${\ds \sup_{t,x}|\pa_{x_1}V(t,x)|}$ because
the right hand side is equal to
$\sin t\pa_{x_1}{\cal U}_0^\ast V(t,x-\ep\sin{t}{\bf e}_1/2){\cal U}_0$.
Hence if $|\mu|\|\pa_{x_1} V\|_{L^\infty} < \ep$, then we have
$i[{\cal L}, {\cal A}]^\circ \geq c>0$.
\end{itemize}
Thus the conditions of \cite{M} are satified and we can conclude that
 $\sigma({\cal K})=\sigma_{ac}({\cal K})=\R$ if $|\mu|\|\pa_{x_1}
V\|_{L^\infty} <
\ep$ by Theorem and Proposition II.4 of \cite{M}.
\vfill\eject

\end{document}